\documentclass{epl}
\usepackage{bbm}
\usepackage{epsfig}
\usepackage{graphicx}
\usepackage[centertags]{amsmath}
\usepackage{amssymb}
\usepackage{amsfonts}
\usepackage{psfrag}

\newcommand{\xv}{{\mathbf x}}
\newcommand{\zv}{{\mathbf z}}

\newcommand{\cv}{{\mathbf c}}

\newcommand{\yv}{{\mathbf y}}

\newcommand{\Tr}{{\rm Tr}}

\newcommand{\xh}{{\hat{x}}}

\def\mm#1{{\mathbf #1}}
\def\Qm{{\mm{Q}}}

\def\ll{{{\mathbf l}}}

\def\Lm{{{\mathbf \Lambda}}}

\title{From Vulcanization to Isotropic \\ 
and Nematic Rubber Elasticity}
\shorttitle{From Vulcanization to Rubber Elasticity}

\author{Xiangjun Xing,\inst{1}
Swagatam Mukhopadhyay,\inst{1}
Paul M. Goldbart\inst{1}\\
\and 
Annette Zippelius\inst{2}}
\shortauthor{X. Xing \etal}

\institute{
\inst{1}Department of Physics, 
University of Illinois at Urbana-Champaign,\\
1110 West Green Street, Urbana, Illinois 61801-3080, U.S.A.\\
\inst{2}Institut f\"ur Theoretische Physik,
  Georg-August-Universit\"at G\"ottingen,\\
  Friedrich-Hund-Platz~1, 37073 G\"ottingen, Germany}

  \date{\today} 


\begin{document}

\maketitle

\begin{abstract}
A Landau theory is constructed for the vulcanization transition 
in cross-linked polymer systems with spontaneous nematic ordering.  
The neo-classical theory of the elasticity of nematic elastomers 
is derived via the minimization of this Landau free energy; this 
neo-classical theory contains the classical theory of rubber 
elasticity as its isotropic limit.  Our work not only reveals 
the statistical-mechanical roots of these elasticity theories, 
but also demonstrates that they are applicable to a wide class 
of random solids.  It also constitutes a starting-point for the 
investigation of sample-to-sample fluctuations in various forms 
of vulcanized matter.

\end{abstract}



The classical theory of rubber elasticity~\cite{ref:Treloar} has been 
remarkably successful.  A blend of phenomenology and molecular-level 
reasoning, it is based on a few simple assumptions, and bears great 
predictive and descriptive power.  By modeling rubbery materials 
(i.e.~elastomers) as incompressible networks of entropic Gaussian 
chains, it gives their elastic free energy density $f$ at temperature 
$T$ as 
\begin{equation}
  f = \frac{\mu}{2}\,{\rm Tr}\,\Lm^{\rm T}\Lm, 
\label{rubber-classical}
\end{equation}
for all uniform deformations ${\bf r}\to\Lm\cdot{\bf r}$
that conserve the volume (i.e.~${\rm det}\,\Lm \equiv 1$). 
For most rubber-like materials the assumption of incompressibility 
is well satisfied.\footnote{This assumption may break down for swollen
rubbers and gels.  In this case, a finite-compressibility version of 
Eq.~(\ref{rubber-classical}) remains valid.  In fact, this is what we
will derive in the paper.}  
The shear modulus $\mu$ is given by $n_{\rm c}\, T$, where the constant 
$n_{\rm c}$ is usually referred as \lq\lq the density of effective chains 
in the network\rlap.\rq\rq\thinspace\  
The classical theory [i.e.~Eq.~(\ref{rubber-classical}) and associated 
arguments] explain many essential features of rubbery materials, such as 
their stress-strain curves (at least for deformations that are not too 
large), and the striking temperature dependence of their shear moduli, 
as well as their strain-birefringence (i.e.~the {\em stress-optical} effect). 

There are several important issues unresolved by the classical theory.  
First, for a given cross-link density, the shear modulus is not calculated 
within the theory.\footnote{We remind the reader that the classical theory 
was developed {\em before} percolation theory was.  Therefore, how an 
infinite network emerges during the random cross-linking process was not 
understood.  In fact, near the vulcanization point, the \lq\lq effective 
chains\rq\rq\ of the classical theory bear little resemblance to the 
original polymer chains before cross-linking.} 
Second, in the intermediate-strain range there is universal and 
significant downward deviation of the experimental stress-strain curve, 
compared with the theoretical prediction.
Finally, the issue of polymer entanglement is not addressed  
by the classical theory.  

Subsequent efforts to improve the classical theory of rubber elasticity have 
focused on various directions (for one overview see Ref.~\cite{GGprime}).  
The non-Gaussian nature of the chain statistics, due to the finite 
extensibility of the polymers, has been taken into account, 
and explains the large-deformation behavior of the stress-strain curve.
Purely mathematical modeling, as in the theories due to Mooney, Rivlin 
and others (see, e.g., Ref.~\cite{ref:Treloar}), also provided useful insight. 
At the microscopic level, the effects of chain entanglement have long been 
emphasized and modeled via various approximation schemes, most notably by 
the Edwards tube model~\cite{Edwards-tube} and its derivatives, although 
results from these models are often either inconclusive or contradictory.  
It seems fair to say that none of these efforts is as successful as the 
classical theory, either in terms of simplicity of assumptions, or in 
terms of broad descriptive power. 

Recently, a simple anisotropic generalization of the classical 
model~\cite{ref:WT-review,Warner-Terentjev}, known as the neo-classical 
model, was constructed to describe the highly unusual elasticity of 
nematic elastomers, i.e., rubbery materials with (spontaneously) broken 
rotational symmetry, and has done so with  great success.  
In the presence of nematic order, the ``step-length tensor'' $\ll$
characterizing the conformations of the polymer chains is anisotropic.  
Then, according to the neo-classical model, the elastic free energy 
of a nematic elastomer with deformation $\Lambda$ is given by 
\begin{equation}
  f  = \frac{\mu}{2}\, \Tr \, \ll_0\,\Lm^{\rm T}\,\ll^{-1}\,\Lm\,,
        \label{Warner-model}
\end{equation}
where $\ll_0$ and $\ll$ are the (in general anisotropic) step-length tensors 
in the initial and deformed states, and are functions of the nematic order 
parameter tensors $\Qm^0$ and $\Qm$ in these two states, respectively. 
A remarkable feature of Eq.~(\ref{Warner-model}) is that for a given $\ll_0$ 
and $\ll$ there exists a continuous manifold of deformations $\Lm$ that cost 
{\em zero} elastic free energy.  These novel soft 
modes~\cite{ref:Lubensky,ref:Olmsted}, as well as the thermal and quenched 
fluctuations associated with them, have been the focus of intensive study, 
both experimentally (see, e.g., Ref.~\cite{Warner-Terentjev}) and theoretically~\cite{Warner-Terentjev,ref:OS-TCL,ref:XX-LR}. 
        
In a classic paper, Deam and Edward~\cite{Deam-Edwards} initiated a 
statistical-mechanical approach to the study of rubber elasticity 
that incorporates both thermal and quenched fluctuations, along with 
repulsive interactions.   This replica-based approach, which has been 
called \lq\lq vulcanization theory\rq\rq\ (VT)
~\cite{vulcan_Goldbart, vulcan_trieste}, has been explored in great detail.  
Progresses in this direction includes, {\it inter alia\/}: 
(i)~the calculation of the mean-field order parameter~\cite{vulcan_Goldbart}; 
(ii)~the derivation of a universal Landau theory~\cite{ref:Peng+Goldbart2000};
(iii)~the development of connections with percolation 
theory~\cite{ref:Peng+Goldbart+McKane2001}; and 
(iv)~the analysis of critical scaling for shear modulus~\cite{shear-modulus-xing}. 
A main virtue of VT is that it follows the Landau paradigm of modern 
condensed matter physics, inasmuch as it concentrates on order parameters, 
symmetries and length-scales.  In particular, because of its independence 
on microscopic details, we view the Landau theory of the vulcanization 
transition as the right theory if one wishes to address the long length-scale 
physics of rubbery materials, especially near the vulcanization transition.          

The aim of the present work is to establish connections between VT and 
the classical---as well as the neo-classical---elasticity of isotropic 
and nematic elastomers.  We begin this task by generalizing the Landau 
theory for the vulcanization transition to systems with spontaneous nematic 
ordering.   As a saddle-point approximation to this theory, we derive the 
neo-classical model~(\ref{Warner-model}) for the elasticity of nematic 
elastomers.  In the isotropic limit, this result reduces to the classical 
theory of rubber elasticity~(\ref{rubber-classical}).  Our work not only 
reveals the statistical-mechanical roots of these elasticity theories, but 
also demonstrates that they are applicable to a wide class of random solids.  
It also constitutes a starting-point for the investigation of 
sample-to-sample fluctuations in various forms of vulcanized matter.  

We begin with the real-space version of the order parameter for the 
replica field theory of the VT~\cite{ref:Peng+Goldbart2000, 
vulcan_Goldbart, shear-modulus-xing,vulcan_unpublished} in $d$ dimensions, 
which is a function of $(1+n)$ $d$-vectors $\xh=(\xv^0,\ldots,\xv^n)$:
\begin{equation}
\Omega(\xh) = \Omega(\xv^0,\ldots, \xv^n) 
                = \sum_{j=1}^N 
			\Big\langle
                        \prod_{\alpha=0}^{n}
			\delta(\xv^{\alpha}-\cv_j^{\alpha})
			\Big\rangle_{1+n}
                -       \frac{N}{V_{0}V^{n}}.
\end{equation}
Here, $\cv_j^{\alpha}$ (with $\alpha = 0,1,\ldots,n$) are the $1+n$ 
replicas of the position $d$-vectors of the $N$ monomers (with 
$j=1,\ldots,N$) that comprise the system.  
$V_{0}$ is the volume of the system in the preparation state, and 
$V$     is the volume of the system in the measurement state 
(which may differ from $V_{0}$).  The brackets 
$\langle\cdots\rangle_{1+n}$ denotes an average over the replica field 
theory given below.  Up to a constant, the order parameter of VT
gives the conditional probability that a monomer found at $\xv^0$ 
at the time of cross-linking is later found at $\{\xv^1,\ldots,\xv^n\}$ 
in $n$ independent measurements after cross-linking.  
The one-replica parts of $\Omega$ (for $\alpha=0,\ldots,n$) 
are defined via
\begin{equation}
\Omega_{\alpha}(\xv^{\alpha})= 
        \int \prod_{\beta (\neq \alpha)} d\xv^{\beta}\,
        \Omega(\xh)
=       \sum_{j =1}^N \Big \langle \delta(\xv^{\alpha} - \cv_j^{\alpha}) 
        \Big\rangle_{1+n} - \frac{N}{{\cal V}_{\alpha}}.  \label{1RPS}
 \end{equation}
where ${\cal V}_{\alpha}\equiv(V_{0},V,\ldots,V)$.  Of these, 
$\Omega_0(\xv^0)$ describes the density fluctuations in the preparation 
ensemble, and 
$\Omega_{\alpha}(\xv^{\alpha})$ (for $\alpha=1,\ldots,n$) describe 
density fluctuations in the measurement ensemble.  

The Landau free energy for VT for isotropic systems is given by
\begin{eqnarray}
H_V[\Omega] 
&=&\int d\xh\left\{
\frac{K_0}{2} (\nabla_0 \Omega)^2
  +\frac{K}{2} \sum_{\alpha = 1}^n
          (\nabla_\alpha \Omega)^2
  +\frac{r}{2} \Omega^2 
  -\frac{v}{3!} \Omega^3\right\}
  \label{Landau-VT}\\ 
  &&
	+\frac{B_0}{2} \int d\xv^0 (\Omega_0)^2
  	+\frac{B}{2} \sum_{\alpha = 1}^n 
        \int d\xv^{\alpha} (\Omega_{\alpha})^2
+\frac{B_0}{2}\frac{N^2}{V_0}
+p_0 \,V_0		
+\frac{n\,B}{2}\frac{N^2}{V }
+n \, p \, V. 
\nonumber
\end{eqnarray}
$(p_0,B_0,K_0)$ and $(p,B,K)$ are, respectively, 
the pressure,  
inverse susceptibility for density fluctuations 
and chain stretchability 
in the preparation and measurement ensembles.  
The chain stretchability is proportional to 
the squared radius of gyration of each constituent polymer chain in the 
isotropic state.  In the absence of any externally imposed deformation, 
the values of $V_0$ and $V$ should be such that they minimize the Landau 
free energy.   For simplicity, we consider the case of equal pressures 
and bulk moduli in the preparation and measurement ensembles 
(i.e.~$p_0 = p$ and $B_0 = B$).   
Near the vulcanization transition, where the order parameter $\Omega$
is small, this leads to $V_0^2 = V^2 = B\,N^2/2p$, in the absence of 
deformation.  The control parameter $r$ triggers the transitions to the 
solid phase when it becomes negative (i.e.~when the density of 
cross-links exceeds some critical value).   This model has been analyzed 
extensively, both within and beyond the mean-field level\footnote{In the 
early literature, the model was usually expressed in momentum space rather 
than of real space.  Furthermore, the one-replica sector of $\Omega$ 
was excluded explicitly, which amounts to setting $B_0$ and $B$ to 
infinity in Eq.~(\ref{Landau-VT}).  The asymmetry between the preparation 
and  measurement ensembles, and its consequences, have been stressed only  recently~\cite{shear-modulus-xing,BWZ-gelation}.}.  
Last but not least, we note that the order parameter $\Omega(\xh)$ of 
this Landau theory is a single-monomer quantity (albeit replicated).  
The original polymer degrees of freedom are integrated out in deriving 
the Landau theory and, consequently, the issue of topological entanglement 
becomes irrelevant in this theory.  It is our belief that inclusion of 
entanglement in the original theory would simply lead to a quantitative 
modification of the parameters in the Landau theory Eq.~(\ref{Landau-VT}), 
not an invalidation of the theory itself.

There are various ways to incorporate nematic ordering into VT.  The 
simplest is to couple the VT order parameter $\Omega$ to the replicated 
symmetric traceless tensor fields $\Qm^{\alpha}$ (with $\alpha = 0,1,\ldots, n$).  
Of these, $\Qm^0$ describes the nematic order in the preparation ensemble, 
whereas $\Qm^{\alpha}$ (for $\alpha = 1,\cdots, n$) describe nematic order 
in the ($n$-fold replicated) measurement ensemble.   The resulting  free energy 
must be invariant under the simultaneous rotation of $\Qm^{\alpha}$and the 
spatial position vectors $\xv^{\alpha}$, independently for each replica $\alpha$. 
The lowest-order coupling (in $\Qm$ and gradients) allowed by symmetry is
\begin{equation}  
H_{\Omega\,Q} =  
\int d\xh \left(
\frac{1}{2}\eta_0\,
Q^0_{ab}\,
\nabla^0_a \Omega\, 
\nabla^0_b \Omega+
\frac{1}{2}\eta\,
\sum_{\alpha=1}^n 
Q^{\alpha}_{ab}\,
\nabla^{\alpha}_a \Omega\, 
\nabla^{\alpha}_b \Omega
\right), 
\label{XQ-coupling}
\end{equation}
where $\nabla_{a}^{\alpha}$ indicates a derivative with respect to the 
$a^{\rm th}$ cartesian component of the $\alpha^{\rm th}$ replicated 
position vector.  The signs of the coupling constants $\eta_0$ and $\eta$ 
depend on details of the chemical structure of nematic polymers under 
consideration.  In a separate publication~\cite{ref:PXMGZ} 
we shall derive Eq.~(\ref{XQ-coupling}) from a lower-level description 

The total free energy should also be augmented by a part that depends 
only on $\Qm^{\alpha}$, and accounts for the nematic interactions between 
neighboring (anisotropic) monomers.  It is known, however, that the nematic 
energy-scale (roughly $k_{\rm B} T$ per monomer) is orders of magnitude 
larger than the energy-scale for the vulcanization transition (roughly 
$k_{\rm B} T$ per chain).  Therefore, we may neglect the feedback of the 
$\Omega$ ordering on the nematic order, and thus may treat $Q^{\alpha}$ 
as given.  More specifically, at the mean-field level, and under the 
assumption of that there is neither replica nor macroscopic translational 
symmetry breaking, we may set both $\Qm^0$ and $\Qm^{\alpha}$ ($\equiv\Qm$ 
for $\alpha=1,\ldots,n$) to be constants, characterizing the uniform 
nematic order in both the preparation and measurement states. 
Additionally, large and positive values for $B_0$ and $B$ guarantee that 
the saddle-point value of $\Omega$ vanishes in the one-replica sector, 
i.e.~Eq.~(\ref{1RPS}) vanishes at the saddle point.  By minimizing the 
total free energy over the VT order parameter $\Omega$, we find the 
saddle-point equation 
\begin{equation}
0=-l^0_{ab}\,\nabla^0_a\,\nabla^0_b\,\bar{\Omega}
-\sum_{\alpha=1}^{n}l_{ab}
\nabla^{\alpha}_a\,\nabla^{\alpha}_b\,\bar{\Omega}
+r\,\bar{\Omega}-\frac{1}{2}v\bar{\Omega}^2\,. 
\label{saddle-equation}
\end{equation}
As they stand, the tensors $\ll_{0}$ and $\ll$ are short-hand for 
\begin{equation}
l^0_{ab}\equiv  K_0\,\delta_{ab}+\eta_0\,Q^0_{ab}\,,
\hspace{3mm}
l_{ab}   \equiv K  \,\delta_{ab}+\eta  \,Q_{ab}\,. 
\label{l-l0}
\end{equation}
However, as we shall see below, they are in fact the effective step-length 
tensors in the preparation and measurement ensembles that appear in the 
neo-classical elastic free energy, Eq.~(\ref{Warner-model}).

In the absence of any externally imposed deformation, the saddle-point 
equation~(\ref{saddle-equation}) is solved by the following 
Ansatz\footnote{A very similar Ansatz in real space has been used in Reference
\cite{BWZ-gelation} for the isotropic case. }:
\begin{subequations}
\begin{eqnarray}
\bar{\Omega}(\xh)&=&
q\int d\zv 
\left\{\int d\tau\,
\frac{p(\tau)}{N(\tau)}\,
\exp\left[
-\frac{\tau}{2} \left(
\yv^0 \cdot \ll_0^{-1} \cdot \yv^0
+\sum_{\alpha = 1}^n 
\yv^{\alpha} \cdot \ll^{-1} \cdot \yv^{\alpha}
\right)
\right] 
-\frac{1}{V_0^{1+n}}\right\},                 
\label{saddle-ansatz}
\\
&&
\yv^0\equiv\xv^0 - \zv,
\hspace{6mm}
\yv^{\alpha}\equiv\xv^{\alpha}-\zv,
\hspace{6mm}
N(\tau)\equiv
\left(\pi/\tau \right)^{\frac{(1+n)d}{2}}
(\det \ll_0)^{\frac{1}{2}}\,(\det\ll)^{\frac{n}{2}}, 
\end{eqnarray}
\end{subequations}
where the $d$-dimensional vector $\zv$ is integrated over the interior
of system $V_0$ in the preparation state.  
Obviously, if both the preparation state and the measurement state are 
isotropic (i.e.~$\Qm^0 = \Qm = {\bf 0}$ and $\ll_0 \propto \ll \propto {\bf I}$, 
the above saddle-point Ansatz reduces to the earlier form appropriate 
to  isotropic systems~\cite{vulcan_Goldbart}, with $p(\tau)$ the so-called 
distribution of inverse square localization lengths.  The interpretation 
of this saddle point is as follows.  A certain fraction of the monomers 
($q$ per unit volume) belong to the infinite cluster (i.e.~the gel 
fraction) and  are localized.  Such a monomer fluctuates around 
the point $\zv$ with Gaussian variance-matrix $\tau^{-1}\ll_0$ in the 
preparation ensemble (i.e.~the $0^{\rm th}$ replica) and fluctuates 
around {\it the same point\/} $\zv$\footnote{This automatically ensures 
that the measurement ensemble has the same volume as the preparation 
ensemble, i.e. $V = V_0$.} with variance-matrix $\tau^{-1}\ll$ 
in the measurement ensemble (i.e.~replicas $1$ to $n$).  From 
Eq.~(\ref{l-l0}) it is easy to see that the role of non-vanishing nematic 
order is to confer anisotropy on these fluctuations.  Finally, because of 
the random nature of elastomers, there is a continuous distribution 
$p(\tau)$ of scales $\tau$.  

In the gel phase (i.e.~$r<0$), we find that Eq.~(\ref{saddle-ansatz}) 
solves Eq.~(\ref{saddle-equation}) provided that  
\begin{equation}
q = 2 |r|, 
\label{saddle-q}
\end{equation}
and that $p(\tau)$ satisfies the following integro-differential equation:
\begin{equation}
 \frac{\tau^2}{2}  p'(\tau)
 =  \left(\frac{|r|}{4\,v}- \tau\right)\,p(\tau) 
 - \frac{|r|}{4\,v} \int_{0}^{\tau} p(\tau')\,p(\tau - \tau')\,d \tau'.
  \label{saddle-tau}
 \end{equation}
Equations~(\ref{saddle-q}) and (\ref{saddle-tau}) are {\it identical\/} 
to those found for the isotropic case of VT~\cite{vulcan_Goldbart}.  The 
stability of this nematic saddle point can also be established in a way 
similar to the isotropic case (see Ref.~\cite{ref:Hessian}). 

\begin{figure}
\begin{center}
\includegraphics[width=6cm]{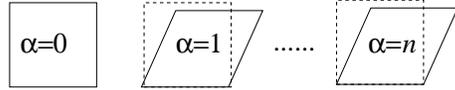}
\caption{An externally imposed deformation changes affinely the 
boundary of the system in the $n$ replicas of the measurement ensemble, 
but not that of the preparation ensemble, i.e. the $0^{th}$ replica. }
\label{shear}
\end{center}
\vspace{-5mm}
\end{figure}

We now come to the main point of this Letter: obtaining the elastic 
free energy of isotropic and nematic random solids.  To do this, we 
shall impose an arbitrary homogeneous deformation of the boundary of 
the system, which we encode in the matrix $\Lm$ and illustrate in 
Fig.~\ref{shear}.  We shall not make any assumptions about how the 
interior of the system changes in response to this deformation.  Our 
aim is to determine the new saddle point $\bar{\Omega}_{\Lm}$ that 
minimises the free energy and is consistent with the deformation of 
the system.  (We explain this consistency further, below.)  We 
proceed by hypothesizing a modification of the original saddle-point 
solution~(\ref{saddle-ansatz}):
\begin{subequations}
\label{modification} 
\begin{eqnarray}
\yv^{\alpha} = \xv^{\alpha} - \zv 
&\longrightarrow& \xv^{\alpha} - \Lm \cdot \zv 
\qquad(\alpha=1,\ldots,n),\\
\frac{1}{V_0^{1+n}} &\longrightarrow&
\frac{1}{V_0\,V^n}.
\end{eqnarray}
\end{subequations}
In general, $\det\Lm$ ($=V/V_0$) may differ from unity, corresponding 
to a change in system volume.  By substituting this modified Ansatz 
into Eq.~(\ref{saddle-equation}), we find that the Ansatz is indeed a 
solution, provided $p(\tau)$ is the {\em same} distribution as 
defined by Eq.~(\ref{saddle-tau}) (in the limit $n \rightarrow 0$).  

The interpretation of this saddle point is as follows.  After the 
deformation, the same fraction of the monomers (i.e.~$q$ per unit volume) 
are localized.  In the preparation ensemble, a localized monomer 
continues to exhibit Gaussian fluctuations around the point $\zv$ with 
an unchanged variance-matrix $\tau^{-1}\ll_0$.  However, in the 
measurement ensemble it fluctuates around the new point $\Lm\cdot\zv$ 
but with the original variance-matrix $\tau^{-1}\ll$.  This implies that 
{\it the average position of each monomer, parametrized by $\zv$, is 
deformed affinely, whereas the fluctuations around $\zv$ remain intact}
\footnote{A similar result was found for isotropic VT 
systems~\cite{elast-Castillo}.}.
Note that $\zv$, as well as $\xv^{0}$, are confined to the volume of 
the preparation ensemble (i.e.~the range over which $\zv$ is 
integrated; see Fig.~\ref{shear}).
Observe, furthermore, that $\bar{\Omega}_{\Lambda}$ vanishes 
whenever $\xv^{\alpha}$ and $\Lm\cdot\xv^{0}$ are widely separated 
(for any $\alpha=1,\ldots,n$).  
It follows that the $\xv^{\alpha}$ are confined to the preparation 
ensemble transformed by the distortion $\Lm$ (i.e.~the measurement 
ensemble), and this establishes that $\Lm$ is the homogeneous 
deformation imposed on the boundary of the system (see Fig.~\ref{shear}).
\footnote{For careful readers we note that, in the isotropic limit, 
this deformed saddle-point solution is slightly different from the 
types of Goldstone fluctuations studied in Ref.~\cite{elast-swagatam},
partly due to different parameterizations.  A  detailed discussion of 
the connections and differences between these two analyses will be 
presented elsewhere~\cite{ref:PXMGZ}.}\thinspace\ 
As we shall show in a forthcoming paper~\cite{ref:stress_fluct}, the 
affine character of the deformation is destroyed by the fluctuations of 
nematic order $\Qm^0$ in the preparation ensemble that become frozen 
in at the time of cross-linking.  

We now calculate the elastic free energy density of nematic elastomers 
$f_{\mbox{{\small el}}}(\Lm)$ at the mean-field level.  To do this, 
we insert the deformed saddle point~(\ref{saddle-ansatz}), modified 
according to Eqs.~(\ref{modification}), into the total free energy 
density~(\ref{Landau-VT}), and subtract its value for the undeformed 
($\Lm={\bf I}$)saddle point.  Then, dividing appropriately by $n$ 
(recall that there are $n$ replicas of the measurement ensemble), 
and taking the replica limit $n \rightarrow 0$, we find 
\begin{subequations}
\label{results}
\begin{eqnarray}
f_{\mbox{{\small el}}}(\Lm) &=& 
\lim_{n \rightarrow 0}\frac{1}{n}  
\left(H[\bar{\Omega}_{\Lm}]-
      H[\bar{\Omega}]\right)
\nonumber\\
&=&\mu\,{\rm Tr} \, \ll_0\,\Lm^{\rm T}\,\ll^{-1}\Lm 
+\frac{1}{2\,\det\Lm} \, \tilde{B}(\det\Lm - 1)^2,
\label{elast-energy}\\
\mu &\equiv& \frac{4}{3} |r|^3\,, 
\hspace{7mm}
\tilde{B}\equiv B_{0}\,\rho_0^2\, .
\label{shear-bulk-modulus}
\end{eqnarray}
\end{subequations}
It is clear that the first term in Eq.~(\ref{elast-energy}) coincides with 
the free energy density of the neo-classical theory of nematic elastomers, 
Eq.~(\ref{Warner-model}); the second term describes the energy cost for 
{\it volume changes\/}, with the bulk modulus $\tilde{B}$ related to the 
parameter $B_{0}$, Eq.~(\ref{shear-bulk-modulus}).  
Consequently, what we have derived, Eqs.~(\ref{results}), 
is the neo-classical elasticity model of nematic elastomers, 
in fact generalized to finite bulk moduli systems.  In the limit 
$\tilde{B}\rightarrow\infty$, the incompressibility constraint 
$\det\Lm \equiv 1$ is restored.   Furthermore, if $\Qm^0 =\Qm = {\bf 0}$ 
we have $\ll_0 \propto\ll\propto{\bf I}$, and our result trivially 
reduces to the classical theory of rubber elasticity, 
Eq.~(\ref{rubber-classical}).   Finally, that the shear modulus, 
Eq.~(\ref{shear-bulk-modulus}), scales as $|r|^3$ is a mean-field 
result and has also been derived via other methods~\cite{elast-Castillo,
elast-swagatam}.   

We emphasize that Eq.~(\ref{elast-energy}) is derived from the Landau 
theory of VT, which includes the the most relevant contributions.   
Therefore it is {\em independent\/} of short-distance details, and thus 
provides a {\em universal\/} mean-field description for the elasticity 
of {\em all forms of vulcanized matter near the vulcanization point\/}, 
provided that the corresponding transition is described by the Landau 
theory.  This observation may explain, in part, the huge success of the 
classical theory of rubber elasticity, Eq.~(\ref{rubber-classical}), and 
its anisotropic generalization, the neo-classical theory, 
Eq.~(\ref{Warner-model}).

The present work also constitutes a starting point for studying 
spatial fluctuations, both thermal and quenched, in vulcanized matter 
of various forms~\cite{Wald-AZ-PMG}.  Sufficiently close to the 
vulcanization point, critical fluctuations of the VT order parameter
field $\Omega$ become important, and change qualitatively the scaling 
of $\mu$.  (This issue was recently addressed; see   
Ref.~\cite{shear-modulus-xing}.)\thinspace\  Nevertheless, 
the {\it form\/} of the elasticity theory~(\ref{Warner-model}) 
continues to hold, upon the incorporation of critical fluctuations.
On the other hand, fluctuations of $\Qm^0$ in the preparation 
ensemble---which may be strong at short length-scales even deep in 
the isotropic phase---would provide a source 
for quenched random stresses in the random solid state, and these 
would couple directly to strain.  Consequently, quenched-in 
fluctuations in $\Qm^0$, and their alignment by external stress, 
may change the elasticity qualitatively, and are likely to be 
responsible for the universal deviation of the stress-strain curves
from the classical theory, observed for half a century.  This issue will 
be explored in forthcoming work~\cite{ref:stress_fluct}.  

\acknowledgments
We thank T.~C.~Lubensky for stimulating discussions.  
This work was supported in part by
grants NSF DMR02-05858 (XX, SM) and 
DOE DEFG02-91ER45439   (XX, PMG), and 
the Deutsche Forschungsgemeinschaft through
SFB~602 (AZ) and Grant No.~Zi~209/6-1 (AZ).





\end{document}